# The Role of Quasi-identifiers in $k$-Anonymity Revisited[*]

Claudio Bettini, X. Sean Wang, Sushil Jajodia [†]

Technical Report RT-11-06
DICo - University of Milan, Italy
July 2006

## 1 Introduction

The concept of *k-anonymity*, used in the recent literature (e.g., [10, 11, 7, 5, 1]) to formally evaluate the privacy preservation of published tables, was introduced in the seminal papers of Samarati and Sweeney [10, 11] based on the notion of *quasi-identifiers* (or QI for short). The process of obtaining $k$-anonymity for a given private table is first to recognize the QIs in the table, and then to anonymize the QI values, the latter being called *k-anonymization*. While k-anonymization is usually rigorously validated by the authors, the definition of QI remains mostly informal, and different authors seem to have different interpretations of the concept of QI.

The purpose of this short note is to provide a formal underpinning of QI and examine the correctness and incorrectness of various interpretations of QI in our formal framework. We observe that in cases where the concept has been used correctly, its application has been conservative; this note provides a formal understanding of the conservative nature in such cases.

The notion of QI was perhaps first introduced by Dalenius in [3] to denote a set of attribute *values* in census records that may be used to re-identify a single or a group of individuals. To Dalenius, the case of multiple individuals being identified is potentially dangerous because of collusion. In [10, 11], the notion of QI is extended to a set of *attributes* whose (combined) values may be used to re-identify the individuals of the released information by using "external" sources. Hence, the appearance of QI attribute values in a published database

[*]**This manuscript has been submitted for publication.**

[†]Claudio Bettini is with University of Milan, Italy, X. Sean Wang is with the University of Vermont, USA, and Sushil Jajodia is with George Mason University, USA. Part of Bettini's work was performed at the University of Vermont and at George Mason University, and part of Wang's work was done while visiting University of Milan and the Chinese University of Hong Kong. The authors acknowledge the partial support from NSF with grants 0242237, 0430402, and 0430165, and from MIUR with grant InterLink II04C0EC1D.

table may give out private information and must be carefully controlled. One way to achieve this control is by anonymizing QI attribute values, through a $k$-anonymization process.

The $k$-anonymization process, as first defined in [10, 11], amounts to generalizing the values of the QI in the table so that the set of individuals, who have the same generalized QI attribute value combination, forms an anonymity set of size no less than $k$, following the pioneering work on *anonymity set* by Chaum [2]. (According to a later proposal for terminology [9], "Anonymity is the state of being identifiable within a set of subjects, the anonymity set.") The resulting table is said to be $k$-anonymous.

The notion of QI is hence fundamental for $k$-anonymity. However, in [10], QI is only informally described. The paper seems to assume that all attributes that may be available from external sources should be part of QI. Recent papers (e.g., [7, 1]) appear to use a similar informal definition, but with a variety of interpretations (see below).

The only formal definition of QI that we found in the literature appears in [11]. The definition is rather complicated, but from that definition, we understand that a set of attributes $Q_T$ in a table $T$ is a QI if there exists a specific individual $r_i$, such that, only based on the combination of the values for $Q_T$ associated with $r_i$, it is possible to re-identify that specific, *single* individual.

From the above formal definition emerges that what really characterizes a QI is the ability to associate a combination of its values with a single individual. The same notion seems to be captured by Def. 2.1 of [6]. We shall call the QI defined this way 1-QI (the number 1 intuitively indicates the number of individuals identified by the QI). This formal definition seems to deviate from the original idea of Dalenius [3] which gave importance to the identification of *groups* of individuals. Although Dalenius was only concerned about collusion, the identification of groups of individuals is closely related to the anonymity set concept and should not be ignored. This deviation actually leads to incorrectness as we shall show in this short note.

Many studies on $k$-anonymization have since appeared in the literature. However, different authors seem to interpret the concept of QI differently. In addition to the original interpretation of QI as (1) the set of all the attributes that appear in external sources [10], and (2) a set of attributes that we call 1-QI [11], we found the following use of QI in $k$-anonymization: (3) use the minimum QI, i.e., the minimum set of attributes that can be used to re-identify individuals [5, 4], and (4) anonymize the multiple minimum QIs in the same table [12] since the minimum QI is found not unique.

Through a formal study of the notion of QI, we conclude in this short note that the use of QI as in category (1) is correct but conservative, while the use of QI as in the other three categories is incorrect. Hence, the contribution of this short note is: (a) the concept of QI and its role in $k$-anonymity are clarified, and (b) the conservative nature of the techniques in the recent papers is better understood. Point (b) above can further lead to (c) new possibilities for more focused data anonymization to avoid over conservativeness.

The remainder of this short paper is organized as follows. Section 2 gives some preliminary definitions. Section 3 introduces a new formalization of $k$-anonymity, and Section 4 defines the notion of QIs and links the QI with $k$-anonymity. Section 5 shows $k$-anonymity using QI other than all the external attributes is problematic, and Section 6 formalizes in our framework the conservative assumption currently used for $k$-anonymization and provides a proof that the approach is sufficient but not necessary. Section 7 concludes the paper.

## 2 Preliminary definitions

Following the convention of the $k$-anonymity literature, we assume a relational model with the bag semantics (or multiset semantics). We assume the standard bag-semantic definitions of database relation/table, attribute and tuple, as well as the standard bag-semantic definitions of the relational algebra operations. In particular, under the bag semantics, relations allow duplicate tuples and operations keep duplicates [8].

We shall use $T$ (possibly with subscripts) to denote relational tables, $t$ (possibly with subscripts) to denote tuples in tables, and $Attr[T]$ to denote the attribute set of table $T$. We shall also use $A$ and $B$ (possibly with subscripts) to denote both sets of attributes and single attributes as the difference will be clear from the context.

To prevent private information from leaking, the $k$-anonymization approach is to generalize the values in a table. For example, both ZIP codes "22033" and "22035" may be generalized to the value "2203*", an interval value [22000–22099], or a general concept value "Fairfax, Virginia". The idea is that each "generalized" value corresponds to a set of "specific" values, and the user of the table can only tell from the general value that the original value is one of the specific values in the set.

The set of specific values that corresponds to a general value can be formally specified with a decoding function. This decoding function, denoted $Dec()$, maps a value to a non-empty set of values. The domain of $Dec()$ is said to be the *general values*, denoted $D_G$, and the range of $Dec()$ is the non-empty subsets of the *specific values*, denoted $D_S$. As such, all attributes in our relational tables will use the same domain, either $D_G$ (for generalized tables) or $D_S$ (for specific tables). We assume that $D_S$ is a subset of $D_G$ and decoding of a $D_S$ value is the set consisting of the value itself. In addition, we assume that the decoding function is *publicly known* and hence all the privacy protection is from the uncertainty provided by the set of values decoded from a single one.

The decoding function is trivially extended to tuples, by decoding each of the attribute values in a tuple. More specifically, given a tuple $t$ with generalized values on attributes $A_1, \ldots, A_n$, $Dec(t)$ gives the set of tuples $Dec(t[A_1]) \times \cdots \times Dec(t[A_n])$, i.e., the cross product of the decoding of each attribute. In other words, the decoding of a tuple $t$ gives rise to the set of all specific tuples that would be generalized to $t$. The decoding function is similarly extended to tables, yielding a set $Dec(T)$ of tables from a given $T$. Specifically, given a table $T = t_1, \ldots, t_n$, a table $T' = t'_1, \ldots, t'_n$ is in $Dec(T)$ if $t'_i$ is in $Dec(t_i)$ for each

$i = 1, \ldots, n$.

In the $k$-anonymization literature, tables may be generalized using a *local encoding* or *global encoding* [5]. (*Encoding* refers to the process of obtaining the general value from a specific one.) The difference is that in global encoding, the different appearances of a specific value are generalized to the same generalized value, while in local encoding, they may be generalized to different generalized values. The formalization with $Dec()$ function is oblivious to this difference, and is correct in the sense that with either approach, the original table is in $Dec(T)$. The $Dec()$ approach is justified as we are not concerned in this short note with specific anonymization techniques.

## 3 The world and $k$-anonymity

In this section, we formally define the notion of $k$-anonymity without using QIs. We will introduce the QI concept in the next section. The approach is in contrast to defining $k$-anonymity based on the concept of QI as traditionally done. We note that our approach is a logical one since only when we can define $k$-anonymity independently of QI, we may prove the correctness of a particular definition of QI.

### The world

To start with, we model all the external sources that can be used to re-identify individuals as a *world*. A world $W$ conceptually is a blackbox that uses attribute values for re-identification. That is, given a tuple $t$ on some of the attributes of $W$, the world $W$ will give back the set of individuals that have the attribute values given by $t$. Formally,

**Definition 1.** *A* world $W$ *is a pair* $(Attr[W], ReID_W)$, *where* $Attr[W]$ *is a set of attributes, and* $ReID_W$ *is a function that maps the tuples on the schemas that are non-empty subsets of* $Attr[W]$, *with domain values from* $D_S$, *to the finite sets of individuals.*

In other words, given a relation schema $R \subseteq Attr[W]$ and a tuple $t$ on $R$ with values from $D_S$, $ReID_W(t)$ gives the set of individuals that possess the attribute values given in $t$. We say that an individual in $ReID_W(t)$ is an individual *re-identified with $t$ by $W$*, or simply *re-identified with $t$* when $W$ is understood. In this case, we may also say that tuple $t$ re-identifies the individual.

Since the $ReID_W$ function re-identifies individuals with their attribute values, one property we call "supertuple inclusion" should hold. For example, if a person is in the set $P$ of individuals re-identified with ZIP code 22032 together with gender male, then this person should be in the set $P'$ re-identified with ZIP code 22032 alone, i.e., $P \subseteq P'$. On the other hand, if a person is in $P'$, then there must be a value of gender (either male or female) so that the person must be re-identified with ZIP code 22032 and gender male (or female) together. More generally, supertuple inclusion property means that if we add more attributes to a tuple $t$ resulting in a "supertuple", then the set of individuals re-identified will be a subset of those identified with $t$, and at the same time, each individual re-identified with $t$ will be re-identified with a particular supertuple of $t$. Formally, we have:

**Definition 2.** *A world $W = (Attr[W], ReID_W)$ is said to satisfy the super-tuple inclusion property if for each tuple $t$ on attribute set $A \subseteq Attr[W]$ and each attribute set $B$, with $A \subseteq B \subseteq Attr[W]$, there exist a finite number of tuples $t_1, \ldots, t_q$ on $B$ such that (1) $t_i[A] = t[A]$ for each $i = 1, \ldots, q$, and (2) $ReID_W(t) = ReID_W(t_1) \cup \cdots \cup ReID_W(t_q)$.*

In the sequel, we shall assume all the worlds satisfy the supertuple inclusion property.

We also assume that, in the sequel, each world we consider is a *closed world*, in which all the relevant individuals are included. That is, the set of individuals identified by $ReID_W(t)$ consists of *all* the individuals who have the attribute values given by $t$. We shall further motivate this assumption at the the end of this section.

A world is called a *finite world* if $ReID_W$ maps only a finite number of tuples to non-empty sets. In the sequel, we assume all worlds are finite worlds.

In summary, we assume in the sequel all the worlds (1) satisfy the supertuple inclusion property, (2) are closed, and (3) are finite.

The function $ReID_W$ in a world $W$ is naturally extended to a set of tuples.

The above conceptual, blackbox worlds may be concretely represented as finite relations. In particular, a world $W = (Attr[W], ReID_W)$ can be represented as a relation $W$ on $Attr[W]$ with domain $D_S$, having the condition that $W$ includes attributes, such as SSN, that directly point to an individual. In this case, function $ReID_W$ will simply be a selection followed by a projection. For example, if SSN is the attribute to identify individuals, then $ReID_W(t)$ is defined as $\pi_{SSN}\sigma_{R=t}(W)$, where $R$ is the schema for tuple $t$.

In this relational view of $W$, table $W$ may be considered as a universal relation storing for each individual all the associated data that are publicly known. As in previous work on this topic, for the sake of simplicity, we also assume that the information of one individual is contained in at most one tuple of $W$. (We will explain in Section 7 how this assumption can be avoided.) We also assume that one tuple of $W$ contains information of only one individual. Furthermore, we assume there is a public method that links a tuple of $W$ with the individual that the tuple corresponds to. This public method may be as simple as an attribute, such as the social security number, in $W$ that directly points to a particular individual.

For example, $W$ may contain the attributes SSN, Name, Birth Date, Gender, Address, Voting record, etc. Each tuple of $W$ corresponds to one individual pointed by the SSN. Other attributes give the other property values of the individual.

Note that the supertuple inclusion property is automatically satisfied by any relational world.

### $k$-anonymity

In our environment, to provide privacy in a published table is to avoid any attacker from using the world $W$ to re-identify the individuals in the published table. The $k$-anonymity is stronger, namely, it avoids any attacker from using the world $W$ to re-identify the individual to be among less than $k$ individuals. This intuition is captured more formally in Definition 4 below.

In order to simplify notation, in the following we use $PAttr[T]$ to denote the *public* attributes of $T$, formally defined as $Attr[W] \cap Attr[T]$ when the world $W$ is understood.

In the above discussion, the case of 0 individuals re-identified is a special case. This is the case when a tuple $\pi_{PAttr[T]}(T)$ does not re-identify anyone by $W$, it would actually be a mistake since $T$ is supposed to represent information of some individuals and the world is assumed to be closed. If this 0 individuals case happens, it must mean that the closed world we have is not "consistent" with the table in our hand. This observation leads to the following:

**Definition 3.** *Given a table $T$, a world $W$ is said to be* consistent *with $T$ if $|\bigcup_{t' \in Dec(t)} ReID_W(t')| > 0$ for each tuple $t$ in $\pi_{PAttr[T]}(T)$.*

A consistent world for a table $T$ is one that can re-identify all the individuals whose information is represented in $T$. In the sequel, we assume the world is consistent with the tables under discussion. We provide motivation for this assumption at the end of this section when we discuss the closed world assumption.

We are now ready to define $k$-anonymity.

**Definition 4.** *Let $k \geq 2$ be an integer, $W$ a world, and $T$ a table with $PAttr[T] \neq \emptyset$. Then $T$ is said to be $k$-anonymous with respect to $W$ if for each tuple $t$ in $\pi_{PAttr[T]}(T)$, we have $|\bigcup_{t' \in Dec(t)} ReID_W(t')| \geq k$.*

In the above definition, the $Dec()$ function is implicitly assumed as public knowledge, and $\bigcup$ is the set union that removes duplicates. Intuitively, the definition says that $T$ is $k$-anonymous if for each tuple $t$ in $\pi_{PAttr[T]}(T)$, we can find at least $k$ individuals from $W$ having values for attributes $PAttr[T]$ as given by $Dec(t)$.

We note that since external information is considered for re-identification, $k$-anonymity should be formally defined with respect to that information, and not simply on the original private table (which may be conservatively considered as a special case, as explained in Section 6), as done in most previous work.

As an example, assume the table in Figure 1(a) is the world, in which the ID attribute is one that directly connects to actual individuals. Table $T$ in Figure 1(b) is 2-anonymous since each tuple (giving either 20032 or 20033 as the zip code value) will re-identify two individuals through $W$. For table $T'$, the decoding function will map name J* to the set of all names that start with J. Hence, the first and the second tuples of $T'$ will re-identify four individuals while the third tuple of $T'$ will re-identify two individuals. Therefore, table $T'$ is also 2-anonymous.

| ID  | FirstName | ZIP   |
|-----|-----------|-------|
| Id1 | John      | 20033 |
| Id2 | Jeanne    | 20034 |
| Id3 | Jane      | 20033 |
| Id4 | Jane      | 20034 |

(a) The world $W$.

| ZIP   | Disease |
|-------|---------|
| 20033 | D1      |
| 20033 | D2      |
| 20034 | D3      |

(b) A table $T$.

| FirstName | Bonus  |
|-----------|--------|
| J*        | $10K   |
| J*        | $100K  |
| Jane      | $20K   |

(c) Another table $T'$.

Figure 1: The world $W$ and two published tables.

### Anonymity and uncertainty

As mentioned earlier, the notion of $k$-anonymity provides protection by forming an anonymity set of size $k$. This notion should not be confused with protection using uncertainty in terms of private values. For example, for $T'$ in Figure 1(c), even if there is only one Bonus value for Jane (hence there is no uncertainty in terms of private values), since there are two Jane's in the world and attackers will not be sure which Jane gets the bonus, we therefore obtain 2-anonymity for Jane. Hence, uncertainty of private values is not a necessary condition for protecting privacy.

However, in other situations, uncertainty is required. For example, take $T$ in Figure 1(b) and assume that the public knows that the first two tuples are for two different individuals. (In most of the $k$-anonymity literature, different tuples in $T$ are assumed to be for different individuals.) Now if the second tuple had disease $D1$ instead of $D2$, there would not be enough protection via anonymity since there are only two individuals in ZIP 20033 in the world $W$. Indeed, in that case, both individuals Id1 and Id3 would have the same disease. The notions of $l$-diversity [6] and $(\alpha, k)$-anonymity [13] are provided to solve this problem.

Two observations arise from the example in the previous paragraph. Firstly, if we do not assume that the public knows the two tuples are for two different individuals, then there is no privacy leaking since there is no way of telling if any of the two individual has the disease (it could be the same individual diagnosed with the same disease twice). The second is that even if the public knows that the two tuples are indeed for two different individuals, 2-anonymity is still maintained for each tuple since there is no way of knowing which of the two indivuals the tuple corresponds to. The privacy leaking is due to a lack of uncertainty. The reader is referred to [14] for more discussion of uncertainty and anonymity (called indistinguishability in [14]). In this short note, we limit our discussion to $k$-anonymity.

### Practical considerations

As observed in [10, 11], in practice it is very difficult to check $k$-anonymity on external sources, mainly due to the difficulty, if not impossibility, of knowing the closed world $W$ that represents the complete knowledge of the external world. Indeed, it is not what we are proposing to do in this short note from an algorithmic point of view. Instead, we use this formal definition to clarify the role of quasi-identifiers, to give a precise semantics to $k$-anonymity, and to study the conservative nature of generalization algorithms reported in the literature.

On the other hand, from a practical point of view, it is possible that some global constraints exist on the world, and that they could be exploited by $k$-anonymization algorithms. For example, if we know from census data that the combination $\langle$ZIP, Gender$\rangle$ has always no less than 500 individuals, any table $T_S$ with $PAttr[T_S] \subseteq \langle$ZIP, gender$\rangle$ is automatically $k$-anonymous for any $k \leq 500$. Further investigation of such a technique is beyond the scope of this short note.

### More on the closed world assumption

The idea of the closed world assumption is that we define $k$-anonymity based on the theoretically "complete" knowledge of the external world. However, it seems to be common in the literature that anonymity is defined based on the possible knowledge of the attackers. By definition, any knowledge an attacker has may very well be a part of the complete knowledge of a closed world.

The question arises as whether we may define $k$-anonymity based on the partial knowledge that an attacker has of the closed world. Two scenarios may be considered. In the first scenario, the attacker does not know all the individuals that a tuple can re-identify. That is, for example, given a ZIP code 22032, the attacker only knows a subset of the individuals who reside in the area determined by this ZIP code. In this scenario, a tuple in $\pi_{PAttr[T]}(T)$ may re-identify by the attacker a proper subset of the $k$ people that can be re-identified by using the closed world. We should not use such partial knowledge for $k$-anonymity for two reasons. (1) The attacker will gain false information in the sense that he/she thinks that the individual is among fewer than there actually are. (2) If we needed to be concerned with the partial knowledge, then we needed to be concerned with all possible partial knowledge. In a particular partial knowledge, the attacker may always re-identify a single individual with any tuple. Then we would not be able gain $k$-anonymity at all. Due to these two reasons, we should remain in our closed world assumption for this scenario.

The other scenario is that the attacker either knows all the individuals that can be re-identified with a tuple by the closed world, or he/she does not know anyone. For example, given a ZIP code 22032, the attacker either know all the individuals living in ZIP 22032, or he/she does not have any clue who might live in the area. This scenario is easier to deal with by simply removing these tuples in $T$ for consideration. However, in order to be conservative, we probably do not want to do that, and again we come back to the conclusion that we need the closed world assumption.

Finally, the fact that a world is closed does not necessary mean that it has the complete knowledge of all the individuals of the whole universe and all their attributes. We only need the closed world to have the complete knowledge about the attributes and the individuals that $T$ is concerned with. For example, if $T$ only has attributes $A_1, \ldots, A_q$ and only concerns residents in the state of Virginia, then the closed world will only need to have the complete knowledge of these attributes and the Virginia residents.

## 4 Quasi-identifiers and $k$-anonymity

In order to understand the relationship between the notion of QI and $k$-anonymity we formally define QI, or more precisely $k$-QI, where $k \geq 1$ is an integer. We then provide a sufficient and necessary condition for $k$-anonymity based on these notions. Intuitively, a set of attributes is a $k$-QI of a world $W$ if a certain combination of values for these attributes can only be found in no more than $k$ individuals of $W$, i.e., if that combination identifies a group of no more than $k$ individuals.

**Definition 5.** *Given a world $W$ and positive integer $k$, an attribute set $A \subseteq Attr[W]$ is said to be a $k$-QI of $W$ if there exists a tuple $t$ on $A$ such that $0 \neq |ReID_W(t)| \leq k$.*

For example, in the relational world $W$ in Figure 1(a), ZIP is a 2-QI, FirstName is a 1-QI, and ⟨FirstName, ZIP⟩ combination is a 1-QI.

Clearly, each set of attributes $A \subseteq Attr[W]$ is a $k$-QI for some $k$ for a given finite world $W$.

Note that the notion of QI formalized in [11] and informally defined in other works is captured by our definition of 1-QI. Indeed, assume some values of QI uniquely identify individuals using external information. That is, if external information is represented by a world $W$, QI is any set of attributes $A \subseteq Attr[W]$ such that $|ReID_W(t)| = 1$ for at least one tuple $t$ on $A$. It can be easily seen that this is equivalent to the notion of 1-QI of $W$.

**Proposition 1.** *If a set of attributes is a $k$-QI, then it is an $s$-QI for each $s \geq k$.*

Thus, we know that each 1-QI is a $k$-QI for $k \geq 2$. It is clear that the inverse does not hold, i.e., if $k \geq 2$ there exist $k$-QI that are not 1-QI. For example, ZIP in the world $W$ of Figure 1(a) is a 2-QI, but not a 1-QI.

**Definition 6.** *A set $A$ of attributes is said to be a* proper *$k$-QI if it is a $k$-QI but it is not an $s$-QI for any $s < k$.*

The following results directly from the supertuple inclusion property of the worlds:

**Proposition 2.** *If a set $A$ of attributes is a $k$-QI, then any $A' \supseteq A$ is a $k$-QI.*

Note that the special case of Proposition 2 for 1-QI has been independently proved in [5].

The following sufficient condition for $k$-anonymity says that if the full set of attributes appearing in external sources is a proper $s$-QI, then the table is $k$-anonymous for each $k \leq s$.

**Theorem 1.** *A table $T$ is $k$-anonymous with respect to a world $W$ if $PAttr[T]$ is a proper $s$-QI in $W$ with $k \leq s$.*

The above theorem holds because by definition, an attribute set $A \subseteq PAttr[W]$ is a proper $k$-QI if for each tuple $t$ on $A$ either $|ReID_W(t)| = 0$ or $|ReID_W(t)| \geq k$. Hence, if $PAttr[W]$ is a proper $s$-QI we know that for each tuple $t$ on $PAttr[W]$, we have either $|ReID_W(t)| = 0$ or $|ReID_W(t)| \geq s$. As we have always assumed that $W$ is consistent with $T$, we know $|ReID_W(t)| \geq s$. For $k$-anonymity, it is enough that we have $s \geq k$.

By the above theorem, if the general constraints on the external world ensure that $PAttr[T]$ is an $s$-QI with $s > k$, then there is no need to anonymize table $T$ if $k$-anonymity is the goal.

Now we can state the relationship between the $k$-anonymity notion and the $k$-QI notion.

**Theorem 2.** *A table $T$ is $k$-anonymous with respect to a world $W$ if and only if for each $k$-QI $A$ of $W$, with $A \subseteq PAttr[T]$, we have $|\bigcup_{t' \in Dec(t)} ReID_W(t')| \geq k$ for each tuple $t \in \pi_A(T)$.*

*Proof.* The "if" part: Assume there is such a $k$-QI $A$. By Proposition 2, we know $PAttr[T]$ is a $k$-QI. By hypothesis and the definition of $k$-anonymity, we know $T$ is $k$-anonymous. If there is no such $k$-QI, then $PAttr[T]$ must be a proper $s$-QI with $s > k$. In this case, by Theorem 1, $T$ is $k$-anonymous.

The "only if" part: Assume $T$ is $k$-anonymous. By the assumption of that $T$ is properly formed, $k$-anonymity of $T$ leads to $|\bigcup_{t' \in Dec(t)} ReID_W(t')| \geq k$ for each tuple $t \in \pi_{PAttr[T]}(T)$. By the supertuple inclusion property of the world $W$, we know $|\bigcup_{t' \in Dec(t)} ReID_W(t')| \geq k$ for each tuple $t \in \pi_A(T)$ and each attribute set $A \subseteq PAttr[T]$ (and hence for each $A \subseteq PAttr[T]$ that is a $k$-QI). □

From the results of this section, we may have the following observations and conclusions. Given a table $T$, if any subset $A$ of $PAttr[T]$ is a $k$-QI, then $PAttr[T]$ itself is $k$-QI. Hence, we need to make sure that the values on $PAttr[T]$, not just a proper subset of $PAttr[T]$, are general enough to gain $k$-anonymity. On the other hand, if we have values on $PAttr[T]$ general enough to have $k$-anonymity, then the values on any proper subset of $PAttr[T]$ will also be general enough due to the supertuple inclusion property. Therefore, for $k$-anonymization, we should only be concerned with the attribute set $PAttr[T]$, not any proper subset of it. In the next section, we shall show, in fact, limiting the consideration to any or all proper subsets of $PAttr[T]$ will lead to privacy leaking.

## 5 Incorrect uses of QI in $k$-anonymization

As mentioned in the introduction, $k$-anonymity in a published table can be obtained by generalizing the values of QI in the table. This process is called *k-anonymization*. As mentioned in the introduction, at least four different uses of QI in $k$-anonymization have appeared in the literature. In this section, we point out the incorrectness of cases (2)–(4). We defer the study of case (1) to the next section.

### 5.1 Use 1-QI only

Firstly, we note that the use of 1-QI (e.g., the QI as defined in [11] and [6]) instead of $k$-QI in the definition of $k$-anonymity can lead to incorrect results. Indeed, accordingly to the current anonymization techniques, if an attribute is not in any QI, then the attribute is not considered for $k$-anonymity or $k$-anonymization (see Def. 2.2 in [6]).

However, if QI is taken as 1-QI as done in [11, 6], it is a mistake.

Consider the table $T$ in Figure 1(b) for 3-anonymity. The public attribute of $T$ is ZIP only, which is *not* a QI (or 1-QI). If we only consider 1-QI for table $T$, then we may incorrectly conclude that the table does not need any generalization (on ZIP values) in order to protect privacy. However, we know $T$ is not 3-anonymous (but is 2-anonymous) against $W$ in the same figure. In order to achieve 3-anonymity, we will need to generalize the ZIP values in $T$.

Therefore, the *k-anonymity requirements* based only on 1-QI fail to protect the anonymity of data when $k \geq 2$. We can correct this problem by considering all $k$-QIs, not just 1-QIs.

### 5.2 Use a subset of *PAttr*[$T$]

The public attributes of a table is given by *PAttr*[$T$]. A few papers seem to imply that only a subset of *PAttr*[$T$] needs to be considered. For example, [5, 4] define QI as the *minimum* subset of *PAttr*[$T$] that can be used to identify individuals, and [12] proposes to generalize all such minimum QIs. Even if we take QI as $k$-QI, the use of the minimum subset is incorrect. We have the following important result.

**Theorem 3.** *Given an arbitrary $T$, an integer $k \geq 2$, and a world $W$, the fact that $\pi_B(T)$ is $k$-anonymous for each proper subset $B$ of PAttr[$T$] does not imply that $T$ is $k$-anonymous.*

*Proof.* We prove the statement by showing that there exist a table $T$ and a world $W$, accompanied by the decoding function $Dec()$, with $Attr[W] = PAttr[T] \cup \{ID\}$, such that $T$ is not $k$-anonymous in $W$ but each projection on a proper subset of *PAttr*[$T$] is $k$-anonymous in $W$. Furthermore, in this world $W$, each subset $A$ of $Attr[W]$ is a 1-QI.

Let *PAttr*[$T$]= $A_1, \ldots, A_n$ and $Attr[W] = PAttr[T] \cup \{ID\}$. For each $i = 1, \ldots, n$, let $Dom(A_i) = \{a_{i1}, \ldots, a_{ik}\}$. Now let $T = Dom(A_1) \times \cdots \times Dom(A_n)$.

The number of tuples in $T$ is $k^n$, and assume we give each tuple a unique ID value from the set $\{1, \ldots, k^n\}$. For each tuple $(a_{1,i_1}, \ldots, a_{n,i_n})$ with tuple ID $r$, we generate the tuple $(a_{1,i_1,r}, a_{n,i_n,r}, r)$ for $W$ and let $W$ consists of all such tuples (and thus it has $k^n$ tuples). We assume that the decoding function works as follows: $Dec(a_{j,i}) = \{a_{j,i,r} | r = 1, \ldots, k^n\}$.

It is clear that in $W$ as constructed above, each subset of $Attr[W]$ is a 1-QI since each value only appears once. We now show that $T$ is not $k$-anonymous while $\pi_V(T)$ is $k$-anonymous for each proper subset $V$ of $PAttr[T]$, and thus proving the proposition.

We first show that $T$ is not $k$-anonymous. Pick an arbitrary $t = (a_{1,i_1}, \ldots, a_{n,i_n})$ in $T$, and assume its ID is $r$. Then $(a_{1,i_1,r}, \ldots, a_{n,i_n,r})$ appears in $\pi_{PAttr[T]}(W)$. By construction of $W$, there are no other tuples of the form $(a_{1,i_1,r'}, \ldots, a_{n,i_n,r'})$ in $W$ with $r' \neq r$. Hence, $T$ is not $k$-anonymous.

Now consider a proper subset $B$ of $PAttr[T]$ with $B = A_1, \ldots, A_p$ and $p < n$. Note that this represents an arbitrary subset due to the symmetry of the attributes in $T$. Take a tuple $t = (a_{1,i_1}, \ldots, a_{p,i_p}) \in \pi_B(T)$. Because of the construction of $W$, we have $(a_{1,i_1,r}, \ldots, a_{p,i_p,r})$ in $\pi_B(W)$ for $k^{n-p}$ different $r$ values since $t$ appears in $k^{n-p}$ number of tuples in $T$. It follows $\pi_B(T)$ is $k$-anonymous since $n > p$ and $(a_{1,i_1,r}, \ldots, a_{p,i_p,r})$ is in $Dec(t)$ for $k^{n-p}$ different $r$ values. □

By Theorem 3, we understand that we cannot simply apply generalization techniques on a proper subset of attributes of $PAttr[T]$. As an example, consider table $T$ and its generalized version $T'$ in Figure 2. Attribute ID is a 1-QI, while ZIP is not a 1-QI (however the combination of ID and ZIP is). To generalize the minimum 1-QI, we would probably generalize table $T$ to $T'$ to make sure there are two appearance for each (generalized) ID value. However, it is clear that $T'$ does not provide 2-anonymity in the world $W$ given in the same figure.

| ID  | Name   | ZIP   |
|-----|--------|-------|
| Id1 | John   | 20033 |
| Id2 | Jeanne | 20034 |
| Id3 | Jane   | 20033 |
| Id4 | Jane   | 20034 |

| ID  | ZIP   | Disease |
|-----|-------|---------|
| Id1 | 20033 | D1      |
| Id2 | 20034 | D2      |
| Id3 | 20033 | D3      |
| Id4 | 20034 | D4      |

(a) The world $W$.  (b) Original table $T$.

| ID        | ZIP   | Disease |
|-----------|-------|---------|
| [Id1–Id2] | 20033 | D1      |
| [Id1–Id2] | 20034 | D2      |
| [Id3–Id4] | 20033 | D3      |
| [Id3–Id4] | 20034 | D4      |

(c) $T'$ with generalized ID values.

Figure 2: Example without proper generalization

## 6 Conservativeness of previous approaches

In practical scenarios, we do not know exactly what the world $W$ is. In such scenarios, we may want to define $k$-anonymity referring to all "possible" worlds, to guarantee "conservative" $k$-anonymity. Indeed, this is the view taken by [10] and other researchers. In this subsection, we provide a formal correctness proof of a common practice of guaranteeing "conservative" $k$-anonymity. (Here, "conservative" means "we would rather err on over protection".)

The common practice we refer to is the following. Given a relational table $T$, assume each tuple contains information about a single, different individual. And assume that the public attributes that can be used to identify the individuals in $T$ are $PAttr[T]$. Then $T$ is $k$-anonymous if for each tuple $t$ in $T$, the value $t[PAttr[T]]$ appears in at least $k$ tuples in $T$. (Note that in the literature, the attributes $PAttr[T]$ above is replaced with the "QI attributes", which would be a mistake if "QI attributes" do not mean $PAttr[T]$ as shown in the previous section.)

In contrast to the definition of $k$-anonymity of this short note, with this common practice, no external world is mentioned. We shall show below that, in fact, this common practice provides $k$-anonymity in a rather "conservative" sense with respect to all "possible" worlds.

We observe that, in contrast to what we have so far, the table $T$ in the common practice has an additional assumption that each tuple of $T$ is for a different individual. Therefore, the requirement of a consistent world for such a table need to be upgraded. Earlier, we only needed a consistent world to be able to re-identify each tuple in $\pi_{PAttr[T]}(T)$ with at least one individual. Here, since each tuple of $T$ is assumed to be for a different individual, a consistent world must be able to re-identify each tuple in $\pi_{PAttr[T]}(T)$ with a different individual.

**Definition 7.** *A world $W$ is said to be* individualized consistent *with a table $T$ with $n$ tuples if there exist $n$ individuals $i_1, \ldots, i_n$ such that there exists $T' = t'_1, \ldots, t'_n$ in $Dec(T)$ satisfying the condition that $i_j$ is in $ReID_W(\pi_{PAttr[T]}(t'_j))$ for each $j = 1, \ldots, n$.*

Intuitively, this means that $T$ could be generalized from a table $T'$ such that each tuple may be used to re-identify a different individual by $W$.

The fact that a world $W$ is individualized consistent with a table $T$ basically confirms the assumption that each tuple of $T$ can indeed re-identify a different individual. All other worlds are going to be "impossible" for table $T$ since the assumption that a different tuple $T$ is for a different individual cannot hold with such worlds. We can now capture the notion of conservative anonymity for such tables.

**Definition 8.** *A table $T$ is said to be* conservatively $k$-anonymous *if it is $k$-anonymous with respect to each $W$ that is individualized consistent with $T$.*

We use the term "conservative" also to indicate the fact that we do not use any knowledge of the world, even if we have any, when $k$-anonymity is consid-ered. In the "practical consideration" part of Section 3, we had an example where knowledge of the world may be used.

We are now ready to show that the common practice described earlier is correct, if $PAttr[T]$ is taken as QI for a given table $T$.

**Theorem 4.** *Let $T$ be a table such that there exists a world that is individualized consistent with $T$. Then $T$ is conservatively $k$-anonymous if for each tuple $t$ in $T$, there exist at least $k-1$ other tuples $t_1, \ldots, t_{k-1}$ in $T$ such that $t_i[PAttr[T]] = t[PAttr[T]]$ for $i = 1, \ldots, k-1$.*

*Proof.* Let $W$ be a world that is consistent with $T$, and $t$ a tuple in $T$. By hypothesis, there exist $k$ tuples $t_1, \ldots, t_k$ in $T$ (may include $t$ itself) such that $t_j[PAttr[T]] = t[PAttr[T]]$, $j = 1, \ldots, k$. By definition of $W$, there exist $k$ individuals $i_1, \ldots, i_k$ such that $i_j$ is in $ReID_W(t'_j)$, where $t'_j$ is in $Dec(t_j)$, for $j = 1, \ldots, k$. Thus, $|\bigcup_{t' \in Dec(t)} ReID_W(t')| \geq |ReID_W(t'_1) \cup \cdots \cup ReID_W(t'_k)| \geq k$. Hence, $T$ is $k$-anonymous wrt $W$. □

Theorem 4 shows that in general, if $PAttr[T]$ is taken as the QI, the common (conservative) process of anonymization appeared in the literature is sufficient under the assumption that we have no knowledge of the world.

The inverse of Theorem 4 does not hold. Indeed, consider $T'$ in Figure 1(c). If the $Dec()$ function is such that $Dec(J*) = \{\text{Jane}\}$ and $Dec(\text{Jane}) = \{\text{Jane}\}$, then it is clear that $T'$ is 3-anonymous with respect to all worlds that are consistent with $T'$ because in any of these worlds, there must be at least 3 individuals with the first name Jane. However, in $T'$ we do not have three tuples with the same First name attribute values.

The above example may be dismissed as using a strange decoding function. However, for any $Dec()$, we can always construct a table $T$ such that the inverse of Theorem 4 does not hold. Formally,

**Theorem 5.** *For any decoding function, the inverse of Theorem 4 does not hold.*

*Proof.* We only sketch the idea of constructing a counterexample table $T$ for the inverse of Theorem 4. First obtain an arbitrary tuple $t$ in $T$ of an arbitrary schema such that $PAttr[T] \neq \emptyset$. We can make such a tuple $t$ to satisfy the conditions $|Dec(t[PAttr[T]])| > 1$ and $t[PAttr[T]]$ is not in $Dec(t[PAttr[T]])$ (otherwise, there is no real generalization going on). Now for each (and every) tuple $t'$ in $Dec(t[PAttr[T]])$, we generate a tuple $t''$ for table $T$ such that $t''[PAttr[T]] = t'[PAttr[T]]$. We duplicate this $t''$ in $T$ for $k$ times. Since we assumed $t[PAttr[T]]$ is not in $Dec(t[PAttr[T]])$, we know that the condition of Theorem 4 is not satisfied for $t$. However, for any world $W$ that is consistent with $T$, there will be at least $k$ individuals in $ReID_W(t[PAttr[T]])$. This is because there must be $Dec(t') = \{t'\}$ for each $t'$ given above, and in this world $W$, $|ReID_W(t')| \geq k$ for each $t'$ in $Dec(t[PAttr[T]])$ by definition of consistent worlds and the fact that $t'$ appears $k$ times in $\pi_{PAttr[T]}(T)$. Hence, tuple $t$ satisfies the $k$-anonymity condition, although $t[PAttr[T]]$ only appears once in $T$. □

In Theorem 3, we showed that $k$-anonymization of any or all proper subsets of $PAttr[T]$ is no guarantee in obtaining $k$-anonymity in $T$. We may extend the result to the conservative case.

**Theorem 6.** *Given an arbitrary $T$ and integer $k \geq 2$, the fact that $\pi_B(T)$ is conservatively $k$-anonymous for each proper subset $B$ of $PAttr[T]$ does not imply that $T$ is conservatively $k$-anonymous.*

*Proof.* We can simply use the table $T$ constructed in the proof of Theorem 3. It is easily seen that $\pi_B(T)$ is conservatively $k$-anonymous for each proper subset $B$ of $\{A_1, \ldots, A_n\}$ due to Theorem 3 and the fact that each tuple $t[B]$ appears for at least $k$ times (as shown in the proof of Theorem 3). To show that $T$ is not conservatively $k$-anonymous we only need to construct one world $W$ that is consistent with $T$ and $T$ is not $k$-anonymous with respect to $W$. In fact, the same $W$ constructed in the proof of Theorem 3 is easily seen consistent with $T$, and we have shown there that $T$ is not $k$-anonymous with respect to that $W$. □

As a final remark of this section, we note that if we do have some knowledge about the world and the $Dec()$ function, we can in some cases do better than this conservative approach. For example, for table $T'$ in Figure 1(c), if we know that $Dec(J*)$ includes Jane, and there are more than 3 Jane's in the world, then $T'$ is 3-anonymous for $T'$. Without such assumptions, we will have to generalize Jane to J* in order to achieve 3-anonymity. The investigation of how to take advantage of such knowledge in anonymization is beyond the scope of this short note.

## 7 Discussion and Conclusion

In summary, we have formally analyzed the notion of quasi-identifier as it is essential to understand the semantics of $k$-anonymity. We have shown that the current formal definitions of QI are not satisfactory and any approach based on these definitions may lead to re-identification (i.e., privacy leakage) as this formally defined QI corresponds to 1-QI as defined in this paper. We also showed the problems with other definitions of QI. We have also formally proved the correctness of using all attributes that appear in external world as QI, and point out precisely what conservative assumptions are made along the way.

We have provided a new formal framework for $k$-anonymity that, by clarifying the role of quasi-identifiers, allows the designers of anonymization techniques to prove the formal properties of their solutions. The presented framework can also serve as the basis for generalization methods with more relaxed, or different assumptions. Indeed, the new notion of $k$-anonymity enables improvements when assumptions can be made on the external information sources, i.e., the world $W$.

Note that all through this short note, we have used "individuals" as the entities whose privacy need to be protected. Obviously, any entities whose privacy need to be protected can be taken as the "individuals", and the notion of $k$-anonymity and $k$-QI should carry over without change.

Finally, it should be mentioned that the assumption of having at most one tuple for each individual in each relational world $W$ can be removed (but each $W$ tuple is still assumed only for one individual) if we assume to have a special attribute $Rid \in Attr[W]$ storing the unique id of the individual for each $W$ tuple. In this case the cardinality of different tuples should be checked on the (set-semantics) projection on this $Rid$ attribute. For example, in Def. 4 the formula $|\bigcup_{t' \in Dec(t)} ReID_W(t')| \geq k$ should be substituted with $|\pi_{Rid} \bigcup_{t' \in Dec(t)} \sigma_{PAttr[T]=t'}(W)| \geq k$. (Here, $|\ |$ counts the number of distinct elements in a bag.)